\documentclass{article}

\usepackage{arxiv}

\usepackage[utf8]{inputenc} % allow utf-8 input
\usepackage[T1]{fontenc}    % use 8-bit T1 fonts
\usepackage{hyperref}       % hyperlinks
\usepackage{url}            % simple URL typesetting
\usepackage{booktabs}       % professional-quality tables
\usepackage{amsfonts}       % blackboard math symbols
\usepackage{nicefrac}       % compact symbols for 1/2, etc.
\usepackage{microtype}      % microtypography
\usepackage{lipsum}
\usepackage{graphicx}
\graphicspath{ {./images/} }

\title{Predicting the ultimate outcome of the COVID-19 outbreak in Italy}

\author{
 Gábor Vattay \\
  Department of Physics of Complex Systems\\
  Eötvös Loránd University\\
  HU-1117 Budapest, Pázmány Péter sétány 1/A \\
  \texttt{vattay@elte.hu} \\
}

\begin{document}
\maketitle
\begin{abstract}
During the COVID-19 outbreak, it is essential to monitor the effectiveness of measures taken by governments on the course of the epidemic. Here we show that there is already a sufficient amount of data collected in Italy to predict the outcome of the process. We show that using the proper metric, the data from Hubei Province and Italy has striking similarity, which enables us to calculate the expected number of confirmed cases and the number of deaths by the end of the process. Our predictions will improve as new data points are generated day by day, which can help to make further public decisions. The method is based on
the data analysis of logistic growth equations describing the process on the macroscopic level. At the time of writing of the first version, the number of fatalities in Italy was expected to be 6000, and the predicted end of the crisis was April 15, 2020. In this new version, we discuss what changed in the two weeks which passed since then. The trend changed drastically on March 17, 2020, when the Italian health system reached its capacity limit. Without this limit, probably 3500 more people would have died. Instead, due to the limitations, 17.000 people are expected to die now, which is a five-fold increase. The predicted end of the crisis now shifted to May 8, 2020.
\end{abstract}

% keywords can be removed

\section{Introduction}
The present study is motivated by the paper \cite{thelancet} that appeared in The Lancet on 13 March 2020, where the authors are calling for trend analysis. At the beginning of the outbreak, Italian data followed closely the exponential growth trend observed in Hubei Province, China. The question is if Italian measures can break the exponential trend? Can they turn the process to a logistic growth with a final saturation value like in China?  

Exponential growth turns into logistic growth when the number of susceptible subjects is limited. If the epidemic is uncontrolled, this limit is defined by the herd immunity, and the vast majority of the population gets infected. Another possibility is when physical contacts in the social network are so severely restricted that the system remains below the epidemic threshold \cite{Vespigniani} so that only finite-sized infected pockets can develop
from infected seeds. This second scenario is in China, where it led to the termination of the epidemic in Hubei Province and only small infected pockets in other parts of China. 

Exponential growth is described by the differential equation
\begin{equation}
    \frac{dN(t)}{dt}=\lambda N(t),
\end{equation}
where $\lambda$ is the rate of growth. If we apply this for the cumulative number of cases $N_n$ on a given day $n$, then the rate of growth can be calculated from the relative growth of the cummulative number of cases
\begin{equation}
    \frac{N_n-N_{n-1}}{N_n} \approx \lambda.
\end{equation}
When the exponential growth has some limit $N_n<N_\infty$, the growth rate depends on the number of cases $\lambda(N)$ and decreases as we approach the
limit. Growth should stop when we reach the limit $\lambda(N_\infty)=0$.
We can expand the growth rate near this point and get
\begin{equation}
    \lambda(N)\approx \lambda'(N_\infty)(N-N_\infty)+\frac{1}{2}\lambda''(N_\infty)(N-N_\infty)^2+...,
\end{equation}
where the first derivative is negatíve $\lambda'(N_\infty)<0$. Introducing
$\lambda_0=-\lambda'(N_\infty)N_\infty$ and $b=\lambda''(N_\infty)N_\infty^2/2$ we can write
$\lambda(N)= \lambda_0(1-N/N_\infty)+b(1-N/N_\infty)^2+...$.
If $b\ll\lambda_0$, the second and higher order terms are negligable and we can
determine $\lambda_0$ and $N_\infty$ from a linear fit
\begin{equation}
    \lambda(N)\approx \lambda_0\left(1-\frac{N}{N_\infty}\right),
\end{equation}
which leads to the logistic growth equation
\begin{equation}
    \frac{dN(t)}{dt}=\lambda_0 N(t)\cdot\left(1-\frac{N(t)}{N_\infty}\right).
\end{equation}
The solution of the logistic equation is the logistic curve
\begin{equation}
    N(t)=\frac{N_\infty}{1+\left[\frac{N_\infty}{N_0}-1\right]e^{-\lambda_0 t}}.
\end{equation}

We can define the end time of the epidemic when the number of cases reaches the fraction $1-p$ of total cases.
The solution of the equation gives  and
\begin{equation}
    (1-p)N_\infty=\frac{N_\infty}{1+\left[\frac{N_\infty}{N_0}-1\right]e^{-\lambda_0 t}},
\end{equation}
where $p\ll 1$ is a very small fraction. The solution for very small $p$ values is then given by

\begin{equation}
    t=\frac{1}{\lambda_0}\log\left[\frac{N_\infty}{N_0}-1\right]-\frac{1}{\lambda_0}\log p \label{time}
\end{equation}

Next, we show that this is an excellent approximation to the actual data, which we
obtained from the JHU site\cite{JHU}.

\section{Deaths as seen on March 17, 2020}

\begin{figure}[htb]
\centering
\includegraphics[width=14cm]{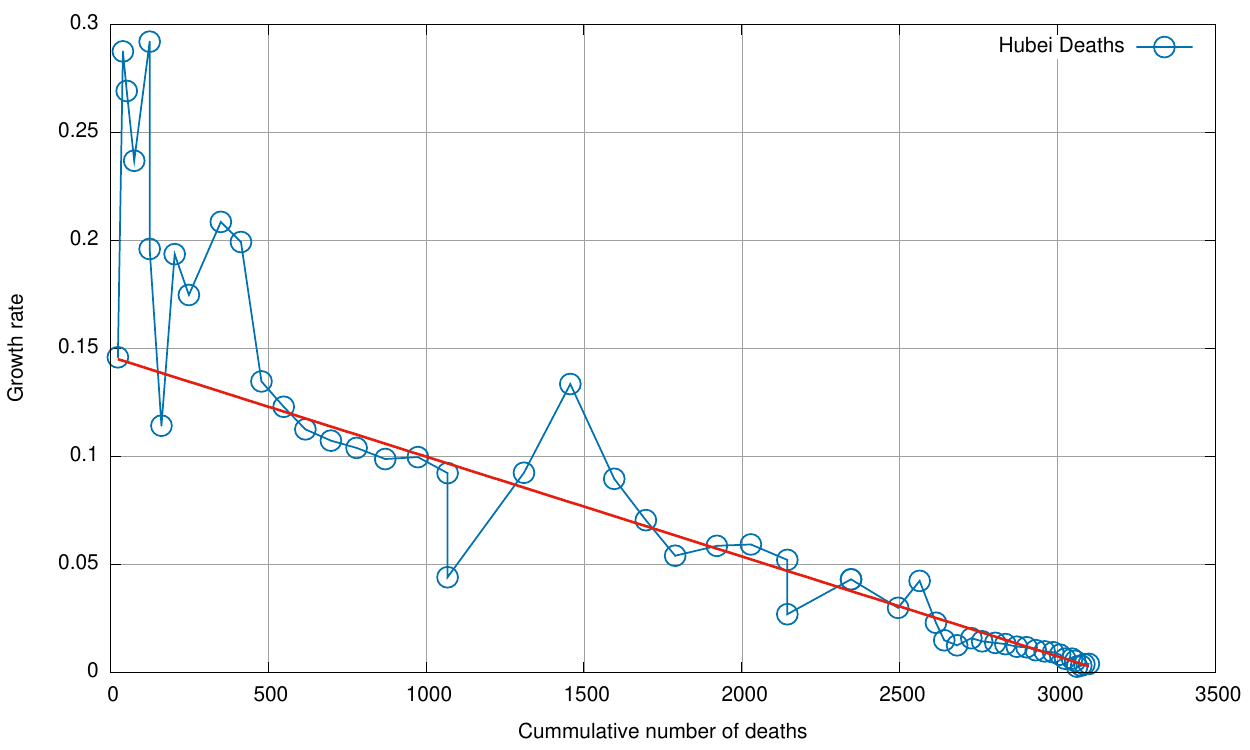}
\caption{Daily growth rate of deaths as a function of the cumulative number
of deaths in Hubei Province, China.
\label{fig1}}
\end{figure}

First, we can do a post-fact analysis of the data from Hubei Province. We have the confirmed cases and deaths data, which both show logistic growth. Both of them have some subjective elements, and never the less the deaths data seems to be more objective.
In Fig.\ref{fig1}, we show the daily growth rate of death cases. Since the reporting of death cases is uneven, deaths are sometimes reported a day later. Therefore we 
calculate the daily rate from the average of two consecutive days
$\lambda_n=0.5\cdot(N_n-N_{n-2})/N_n$.
The first data point is January 24, 2020, three days after the lockdown of Wuhan. The daily growth rate is fluctuating around 25-30\%, which seems to be the "natural" rate of growth in the absence of public actions. Then 14 days after the lockdown at about 500 deaths, the daily growth rate drops to about 10\%. Then a linear trend sets in with some fluctuations due to some errors in the reporting. The slope of the linear fit is $\lambda_0/N_\infty=4.6\pm 0.4 \cdot 10^{-5}$, $\lambda_0\approx 0.146$, and $N_\infty\approx 3158$.

\begin{figure}[htb]
\centering
\includegraphics[width=14cm]{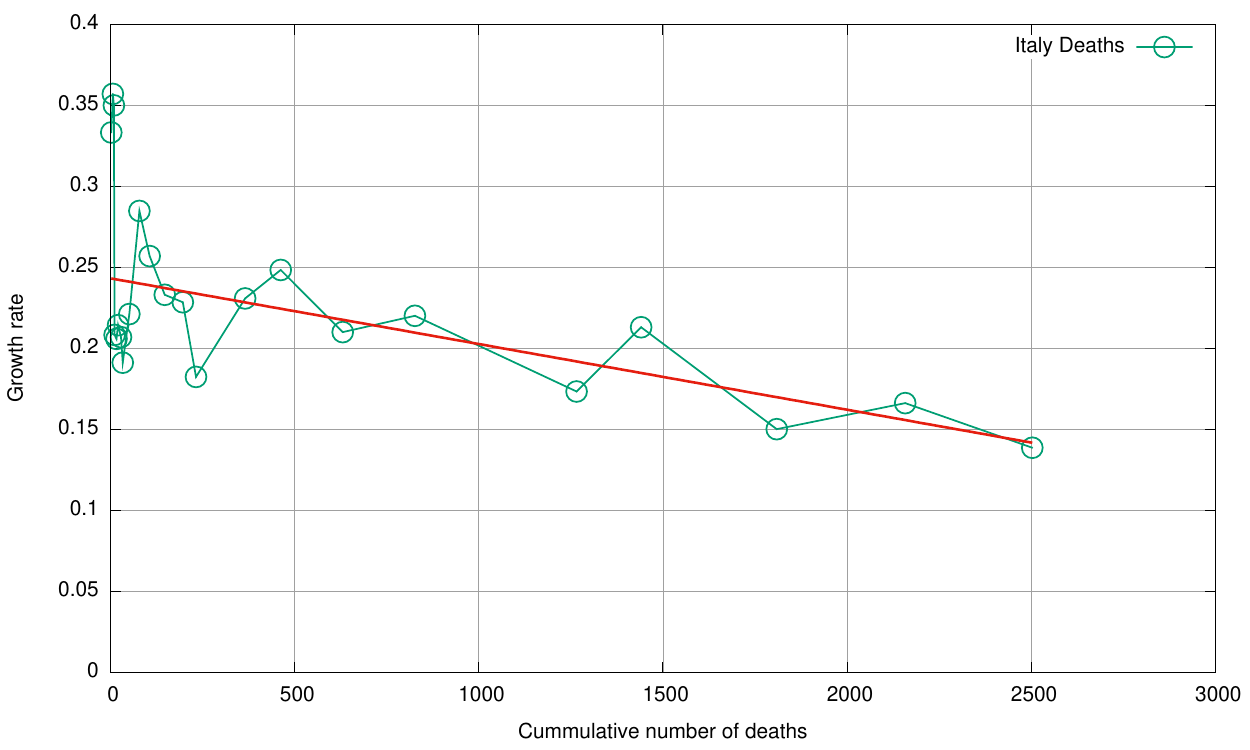}
\caption{Daily growth rate of deaths as a function of the cumulative number
of deaths in Italy.
\label{fig2}}
\end{figure}

We repeat the same analysis for the data from Italy and show the result in Fig.\ref{fig2}. The Italian data starts on February 25, 2020. The beginning
of the data fluctuates around 20-30\% daily growth. The lockdown of Lombardy starts
on March 9, at about 463 deaths, and the linear trend can be observed starting from this date. The
slope of the linear trend is $\lambda_0/N_\infty=4.05\pm 1.13 \cdot 10^{-5}$, $\lambda_0\approx 0.243$, and $N_\infty\approx 5996$.

Remarkably, the slope of the curve is the same in Hubei and Italy within the error of fitting. It seems that the lockdown has a similar effect in
both cases. The daily rate of growth of deaths appears to be falling 4\% for each
1000 new deaths in Italy and the daily growth rate will be reduced from roughly 24\% 
to zero after 6000 deaths. In Hubei the initial daily growth rate of 12\% fell to zero after about 3000 deaths. It seems that this is a universal consequence of lockdown,
which needs further explanation.

\section{Deaths after March 17, 2020}

\begin{figure}[htb]
\centering
\includegraphics[width=14cm]{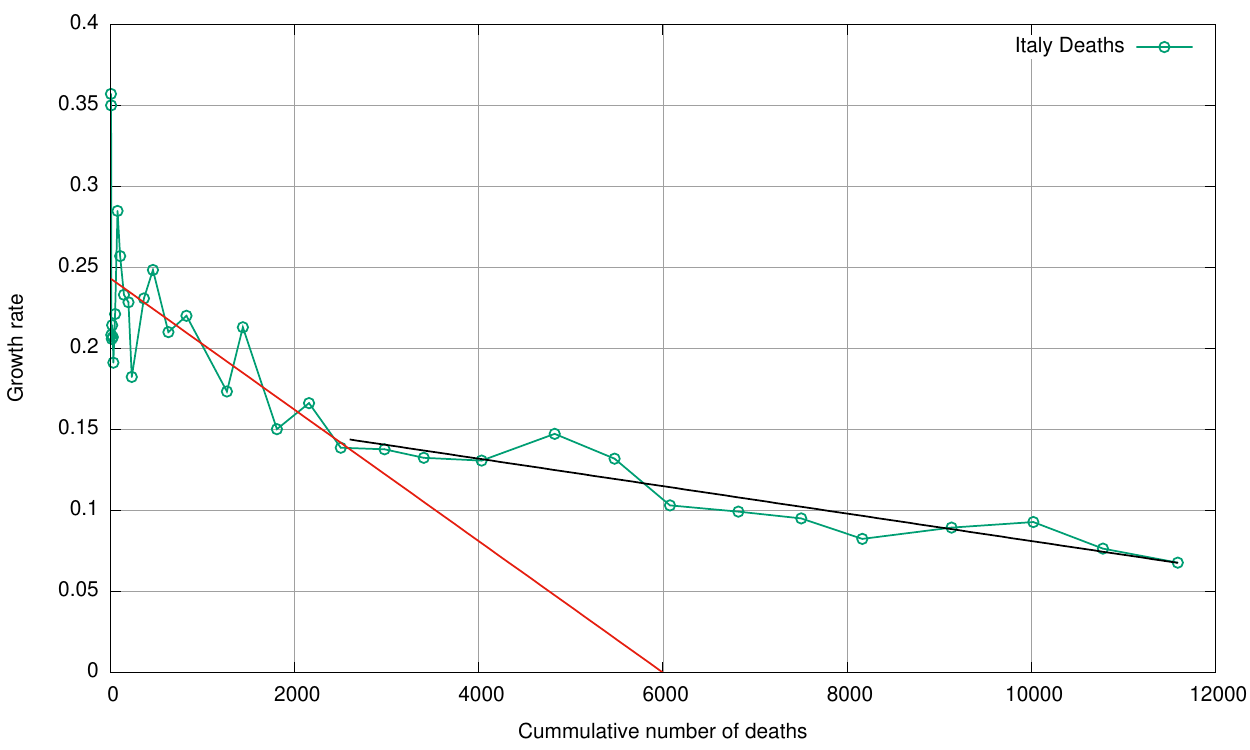}
\caption{Daily growth rate of deaths as a function of the cumulative number
of deaths in Italy. The red line is the same as in Fig.\ref{fig2}, while the black line
is fitted to the data from March 17 to 30, 2020. 
\label{fig3}}
\end{figure}

After releasing the first version of this manuscript, the situation in Italy started to change. In Fig.\ref{fig3}, we show the daily growth rate of fatalities again as a function of the cumulative number of deaths. We keep the original linear fit too, which would lead to about 6000 deaths. Instead, the slope changed sharply on March 17, and
the new, clearly visible trend predicts about $N_\infty=19557\pm 1670$ deaths. 
The new slope is $\lambda_0/N_\infty=8.1 \cdot 10^{-6}$, and the rate of deaths falls only 0.8\% for every 1000 new deaths. The reason for this radical change in the slope is probably not related to the progress of the epidemic. The number of deaths is simply a proxy for the size of the epidemic. It depends on the number of people infected and the
probability of death. We assume that the latter changed drastically when the Italian health system reached its capacity limit on March 17. Without this limit, probably 3500 people would have died. Instead, due to the limitations, 17.000 people are expected to die now, which is a five-fold increase.

\section{Length of the outbreak as seen on March 17, 2020 }

Using the fitted parameters, we attempt to calculate the expected end of the outbreak in Italy. First, we validate the model on the Hubei data. In Fig.\ref{fig1}, the linear regime starts when the number of deaths is $N_0=549$ on February 5. Today, 40 days later, on March 17, the number of deaths is 3111 in Hubei, which is 98.5\% of the expected deaths $N_\infty=3158$ or $p=0.015$. Using these numbers (\ref{time}) yields
$t=39.44$ days, which is a very precise estimate for the length of the process. 
In Italy, the linear regime starts on March 9, and $N_0=463$. Using the same $p=0.015$ and the number of ultimate deaths $N_\infty=5996$
(\ref{time}) yields $t=27.49$ days that is April 6. The $p=0.001$ milestone will be reached around April 15, and the lockdown can probably be partially lifted. 

\section{Length of the outbreak after March 17,2020}

We can repeat the calculation with the new data $N_\infty=19557$ and $N_0=11591$ on
March 30, 2020. For $p=0.015$ the result is 23 days (April 22, 2020) and the $p=0.001$
milestone is reached on May 8, 2020. Hopefully.

I would like to thank the help of István Csabai, Orsolya Pipek, Krisztián Papp, and his team for continuous discussions and sharing the data with me. Our current forecast for the outcome of COVID-19 in various countries is now online\cite{online}. 

\bibliographystyle{unsrt}  


\begin{thebibliography}{1}

\bibitem{thelancet}
Andrea Remuzzi and Giuseppe Remuzzi.
\newblock COVID-19 and Italy: what next?,
\newblock The Lancet 13 March 2020, https://doi.org/10.1016/S0140-6736(20)30627-9

\bibitem{Vespigniani}
Pastor-Satorras, Romualdo and Vespignani, Alessandro.
\newblock Epidemic Spreading in Scale-Free Networks,
\newblock Phys. Rev. Lett. 86, 14, 3200--3203 (2001)
https://link.aps.org/doi/10.1103/PhysRevLett.86.3200

\bibitem{JHU}
https://coronavirus.jhu.edu/

\bibitem{online}
https://kooplex-fiek.elte.hu/notebook/report-pkrisz5-covid19dashmodel---/report/

\end{thebibliography}

\end{document}